\documentclass[aps,prl,twocolumn,showpacs,10pt,preprintnumbers,nofootinbib]{revtex4-1}

\usepackage[margin=7pt,justification=centerlast]{caption}

\usepackage{amsmath}
\usepackage{amsfonts}
\usepackage{amssymb}
\usepackage{graphicx, rotating}
\usepackage{epstopdf}
\usepackage{epsfig}
\usepackage{latexsym}
\usepackage{mathtools}
\usepackage{MnSymbol}
\usepackage{graphicx}
\usepackage{color}
\usepackage[dvipsnames]{xcolor}
\usepackage{amsmath,amssymb}
\usepackage{multirow}
\usepackage{hhline}
\usepackage{array,makecell}
\usepackage[utf8]{inputenc}
\usepackage{arydshln}
\usepackage{ mathrsfs }

\usepackage{lipsum} 

\usepackage{tikz-feynman}
\tikzfeynmanset{warn luatex=false}

\usepackage{slashed}
\usepackage{hyperref}
\hypersetup{colorlinks, citecolor=black, linkcolor=black, urlcolor=bluscuro}
\definecolor{rossos}{cmyk}{0,1,1,0.55}
\definecolor{bluscuro}{rgb}{0.15, 0.2, .85}
\definecolor{bluchiaro}{cmyk}{1,.3,0.,0.1}
\definecolor{verdescuro}{rgb}{0.3,0.8,0.3}

\newcommand{\nn}{\nonumber}

\newcommand{\be}{\begin{equation}}
\newcommand{\ee}{\end{equation}}          
\newcommand{\bea}{\begin{eqnarray}}
\newcommand{\eea}{\end{eqnarray}}
\newcommand{\bc}{\begin{center}}
	\newcommand{\ec}{\end{center}}

\newcommand{\M}{{\cal M}}

\begin{document}

\title{Full Unitarity and the Moments of Scattering Amplitudes}

\preprint{CERN-TH-2022-210}

\author{Marc Riembau}

\affiliation{CERN, Theoretical Physics Department, Geneva CH-1211, Switzerland}

\begin{abstract}
\noindent 
{
We study the impact of full unitarity on the moment structure of forward scattering amplitudes. We introduce the semiarcs, calculable quantities in the EFT dispersively related to both real and imaginary parts of the UV amplitude for a fixed number of subtractions.
It is observed that large hierarchies between consecutive moments are forbidden by unitarity.
Bounds from full unitarity compete with the ones stemming from convexity, and become more important in EFTs where the loop expansion is more important than the derivative expansion.
}

\end{abstract}

\maketitle

\section{I.\quad Introduction}

Our description of relativistic quantum phenomena is based on the framework of effective field theories (EFT), that describe long-distance dynamics of a system.
Effects of the microscopic dynamics are encoded in a set of parameters that, although thought arbitrary from the low-energy, infrared (IR) point of view of the EFT, obey positivity constraints coming from dispersion relations \cite{PhysRevD.31.3027,Ananthanarayan:1994hf,Pennington:1994kc,Comellas:1995hq}.
As shown in \cite{Adams:2006sv}, negativity of certain coefficients of the forward amplitude leads to long distance violations of causality. 
Such constrains on the EFT space of parameters from structural UV consistency of the theory have been extended to an infinite set of two-sided bounds on the Wilson coefficients \cite{deRham:2017avq,Bellazzini:2020cot,Tolley:2020gtv,Caron-Huot:2020cmc,Arkani-Hamed:2020blm}. 
Such positivity bounds have been shown useful to understand generic QFT properties \cite{Nicolis:2009qm,Komargodski:2011vj,Luty:2012ww,CarrilloGonzalez:2022fwg,Creminelli:2022onn,Chowdhury:2021ynh,Alberte:2021dnj,Sinha:2020win,Meltzer:2021bmb,Bellazzini:2021oaj}, gravity \cite{Camanho:2014apa,Bellazzini:2015cra,Cheung:2016yqr,Bellazzini:2017fep,Chen:2019qvr,Bellazzini:2019xts,Bern:2021ppb,Caron-Huot:2021rmr,Haring:2022cyf,Serra:2022pzl} and particle phenomenology \cite{Distler:2006if,Vecchi:2007na,Low:2009di,Bellazzini:2014waa,Englert:2019zmt,Remmen:2019cyz,Zhang:2020jyn,Bonnefoy:2020yee,Chala:2021wpj,Zhang:2021eeo,Gu:2020ldn,Alberte:2020bdz,Henriksson:2021ymi,Davighi:2021osh,Henriksson:2021ymi,Albert:2022oes,Haring:2022sdp,Fernandez:2022kzi}.

Such constraints make only partial use of the unitarity of the S-matrix, in the sense that only the positivity of the imaginary part of the amplitude is invoked. 
Full non-linear constraints are central in the modern incarnation of the S-matrix bootstrap \cite{Paulos:2016fap,Paulos:2016but,Paulos:2017fhb,Homrich:2019cbt}, which has provided a quantitative understanding of a wide range of strongly coupled systems \cite{Guerrieri:2018uew,Cordova:2018uop,Cordova:2019lot,EliasMiro:2019kyf,Guerrieri:2020bto,He:2021eqn,EliasMiro:2021nul}. 
Finding common ground between this numerical approach and the more qualitative but analytic and versatile idea of dispersion relations might bring novel perspectives on EFTs emerging from strongly coupled dynamics \cite{EliasMiro:2022xaa}.
Implications of full unitarity in the EFThedron formulation of the EFT bounds has been explored in \cite{Chiang:2022ltp}.
In this \textit{Letter} we discuss the implications of full unitarity in the UV to the structure of positive moments of the forward amplitude presented in \cite{Bellazzini:2020cot}. We argue that the moment structure is a necessary, but not sufficient consistency condition of the forward scattering amplitude. 

Naively, one might be tempted to think that full unitarity should simply translate into an upper bound on the overall coupling captured by first moment $a_0$ since, after all, \textit{any} positive measure leads to the structure of moments. However, the real part of the amplitude at a given point depends on the convolution of its imaginary part along a certain contour, and therefore full unitarity must set constraints not only to the overall normalization of the measure but also on its \textit{functional form}. 
Ruling out a given measure translates into ruling out a subspace of arcs, and therefore a subspace of EFT parameters.

In the following we shall formalize this simple idea. For this we need to introduce an appropriate contour in the complex plane, that we call \textit{semiarcs}, that allows to dispersively relate EFT-calculable quantities to both real and imaginary parts of the UV amplitude for a fixed number of subtractions.
We remark the central role played by null constraints \cite{Caron-Huot:2020cmc} in order to transfer the unitarity bounds on partial waves to the semiarcs.
In the context of an EFT of a Goldstone boson, we observe how the constraints from full unitarity become more relevant than the ones coming from the semipositiveness of the Hankel matrix of moments (see e.g. \cite{Bellazzini:2020cot,Arkani-Hamed:2020blm}) when the loop corrections are more important than higher orders of the derivative expansion.

\section{II.\quad Semiarcs}

We study the $2\to2$ scattering amplitude of a derivatively coupled massless scalar. We focus on the scattering amplitude in the forward limit, $\mathcal{M}(s) = \lim_{t\to 0}\mathcal{M}(s,t)$.
This choice is made for simplicity and it is expected than similar arguments can be used for $t\neq 0$, see e.g. \cite{Bellazzini:2021oaj}.
As a function of the complex Mandelstam variable $s$, the scattering amplitude $\mathcal{M}(s)$ is assumed to satisfy the following analytic properties \cite{Martin:1969ina}: \textit{i)} Crossing symmetry, $\M(s)=\M(-s)$, together with real analyticity, $\M(s)=\M^*(s^*)$. \textit{ii)} The sole analyticity of $\M(s)$ is the branch cut at physical energies $s>0$, and the corresponding crossing symmetric cut at $s<0$. \textit{iii)} Unitarity of the S-matrix $S=1+i \mathcal{T}$, which implies $2\text{Im}\mathcal{T}>\mathcal{T}^\dagger \mathcal{T}$ for physical energies. In particular, $\text{Im}\M(s)>0$. \textit{iv)} Polynomial boundedness of $\M(s)$, in particular the Froissart bound $\M(s)/s^2\to0$ as $|s|\to\infty$ \cite{PhysRev.123.1053,PhysRev.129.1432,PhysRev.135.B1375}.

These set of properties impose nontrivial constraints on the functional form that the amplitude $\M(s)$ (or $\M(s,t)$, in general) can take. A given representation of the amplitude might obscure certain properties. Dispersion relations allow to relate the amplitude in different kinematic regimes
, imposing nontrivial constraints on the explicit form of a given amplitude. For instance, as we will discuss in detail in the following, unitarity in the UV implies positivity of certain coefficients in the IR, and crossing in the IR implies certain sum rules in the UV partial waves.

\begin{figure}
	\centering
	\includegraphics[width=1\linewidth]{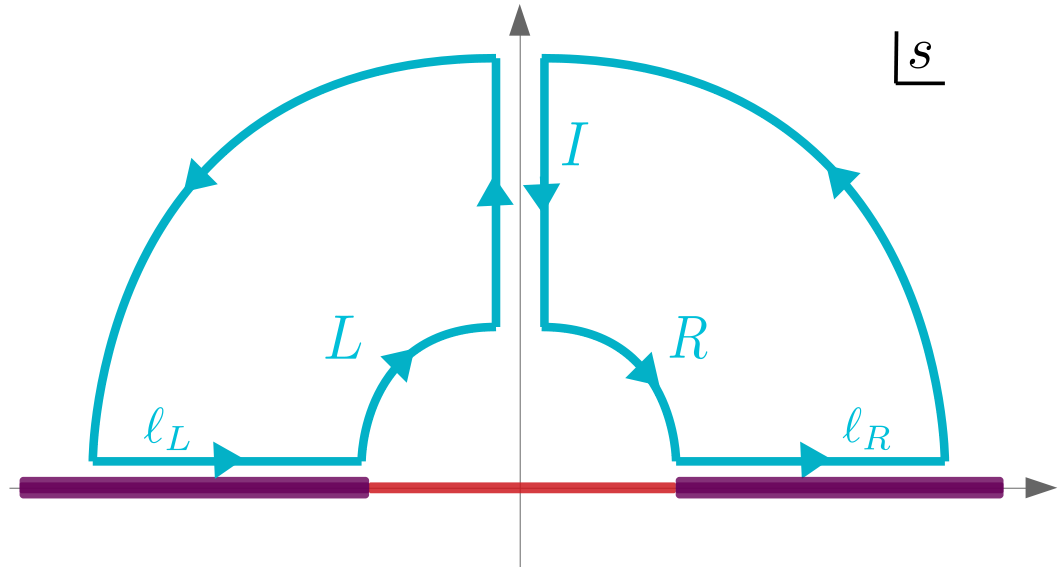}
	\caption{Contours in the complex $s$-plane that define the \textit{semiarcs}. The right contour leads to the relation in Eq.~\ref{eq:rightcontour}, while the left one leads to Eq.~\ref{eq:leftcontour}.}
	\label{fig:lrcontour}
\end{figure}

\smallskip
In order to explicitly reveal such constraints, we consider the integrals of $\M(s)/s^{n+3}$ along the contours on the left and right sides of the upper half plane in Fig.~\ref{fig:lrcontour}.
The right contour gives the relation
\bea\nn
0&=& \left(\int_{i\infty}^{is}-\int_{R}+\int_{s}^\infty\right)\frac{dz}{i\pi} \frac{\M(z)}{z^{n+3}}\\
&\equiv& I_n(s) - a_n^R(s) + A_n(s).
\label{eq:rightcontour}
\eea
Note that we have defined the integral over the semiarc $R$ counterclockwise. We will refer to $a_n^R(s)$ as the \textit{semiarc}. The scale $s$ on which the semiarcs $a_n^R(s)$ depend, being the radius of the semiarc, should be thought as the scale at which the EFT is explored, and therefore smaller than the maximal scale where such description stops being valid.

Along the imaginary axis, one has $z^*=-z$, and therefore crossing together with real analyticity imply that the amplitude is purely real, $\M(z)=\M^*(z)$. This implies that $I_n$ is purely imaginary (real) for $n$ even (odd). For $n$ even, one gets the relations
\bea\label{eq:imarcs}
2\text{Re}(a_n^R) &=& \frac{2}{\pi}\int_{s}^\infty \frac{dz}{z}\frac{\text{Im}\M(z)}{z^{2+n}} \equiv A_n^{\text{Im}}\\
2\text{Im}(a_n^R) &=& - A_n^{\text{Re}} -2i I_n,
\eea
where $A_n^{\text{Re}}$ is defined similarly to $A_n^{\text{Im}}$. The first relation reveals that $2\text{Re}(a_n^R)$ are indeed the arcs of Ref.~\cite{Bellazzini:2020cot}. Positivity of the imaginary part of the amplitude implies the identification of $A_n^{\text{Im}}$ with the moments of a positive distribution, in particular with the Hausdorff moment problem (see e.g. \cite{lasserreJB}). While $A_n^{\text{Im}}$ is a UV quantity, dispersion relations allow to find an IR representation in terms of the semiarc $2\text{Re}(a_n^R)$, which is calculable within the EFT and inherits the moment structure.

The imaginary part of the semiarcs is proportional to the integral of the \textit{real} part of the amplitude, $A_n^{\text{Re}}$. Therefore, if one has some control over the amplitude along the crossing symmetric axis $I_n$, full unitarity in the UV may be dispersively related to constraints on the structure of the semiarcs in the IR.

Before continuing the discussion about $I_n$, we shall comment on the absence of the contour along the negative real axis. In the usual formulation of dispersion relations, the negative axis cut is related to the contribution along the physical axis using crossing and real analyticity. In order to obtain Eq.~\ref{eq:imarcs}, we have explicitly used the same crossing and real analyticity assumptions to conclude that $\M(s)$ is purely real along the crossing symmetric axis. Therefore, both constructions are equivalent. Due to crossing, the left side of the complex plane is redundant and the right contour is enough. Indeed, the left contour leads to the relation
\be
0 = -I_n(s) - a_n^L(s) - (-1)^n A_n^*(s).
\label{eq:leftcontour}
\ee
The object $A_n^*(s)$ is defined like $A_n(s)$, but integrating the complex amplitude $\M^*$ instead, and $a_n^L$ is similar to $a_n^R$, but integrating along the counter clockwise contour $L$. Using the usual properties, it is possible to show that $a_n^L = (-1)^n(a_n^R)^*$. 
Together with Eq.~\ref{eq:rightcontour}, one gets the relations
\bea\nn
a_n\,\equiv\, a_n^R + a_n^L &=& A_n -(-1)^nA_n^*\\
a_n^- \,\equiv\, a_n^R - a_n^L &=& A_n +(-1)^nA_n^* + 2I_n.
\label{eq:sumanddifferencearcs}
\eea
The sum of the semiarcs $a_n$ is of course the arc of Ref.~\cite{Bellazzini:2020cot}, and proportional to the integral of the imaginary (real) part of the amplitude in the UV for $n$ even (odd). The difference of semiarcs $a_n^-$ is instead proportional to the real (imaginary) part of the amplitude in the UV for $n$ even (odd), plus the crossing symmetric axis contribution $I_n$.

\textit{The crossing symmetric integral $I_n$---}.
We still need to determine the contribution from the crossing symmetric axis $I_n$.
While it might be possible to use properties of the amplitude in this region to learn and understand the physical meaning of such contribution \cite{Haring:2022cyf}, we opt in the following for a pragmatic approach.
We will relate the integral $I_n$ to integrals along the contours $L$, $R$ and $\ell_R$ that contain a kernel different than $1/z^n$.
Indeed, $I_n$ can actually be written in terms of the discontinuity along a logarithmic branch cut in the imaginary axis, generated by $\log(is)$. Integrating the quantity $\M(z)\log(iz/s)$ along the right and left contours, one obtains
\bea\nn
I_n =-\frac{1}{2\pi i}\left(\int^{-s}_{-\infty}-\int_{L}-\int_{R}+\int_{s}^\infty\right)\frac{dz}{i\pi} \frac{\M(z)}{z^{n+3}}\log\left(\frac{iz}{s}\right)\\\nn
= -\frac{1}{\pi i} \bigg[ A_n^{\text{log}}-(-1)^nA_n^{*,\text{log}} - \left( a_n^{R,log} +(-1)^n (a_n^{R,\text{log}})^* \right)  \bigg]
\eea
where the label ${}^{\text{log}}$ indicates the presence of a $\log\frac{z}{s}$ in the integrand, as in
\be
a_n^{R,log} =\int_R\frac{dz}{z}\frac{\M(z)}{z^{n+2}}\log\frac{z}{s},
\ee
which is calculable in the EFT. 

For even $n$, the quantity $A_n^{\text{log}}-(-1)^nA_n^{*,\text{log}}$ is equal to $A_n^{\text{Im},\text{log}}$, i.e. the integral over the imaginary part as in Eq.~\ref{eq:imarcs} but weighted by $\log\frac{z}{s}$. Given the positivity of $\text{Im}\M$, it is possible to write $A_n^{\text{Im},\text{log}}$ in terms of a linear combination of $A_{n^\prime}^{\text{Im}}$ just by a change of variables. Indeed, the integral can be mapped to the unit segment $(0,1)$ and the logarithm can be written in terms of polynomials in a way that all integrals can be written as IR arcs. In the following we do not take this route and use a non optimal but very useful and intuitive shortcut. It is indeed convenient to use polynomials in $z^2$ and $1/z^2$ since this gives rise to bounds in terms of higher and lower arcs. From the relation
\be
\frac{1+2\log\frac{z_-}{s}}{2}-\frac{z_-^2}{2z^2}\,<\log\frac{z}{s} \,<\,\frac{-1+2\log\frac{z_+}{s}}{2}+\frac{z^2}{2z_+^2},
\ee
valid for all $z_-,z_+ \geq s$, one obtains the linear bounds
\bea\nn
\frac{1+2\log\frac{z_-}{s}}{2}a_n&-&\frac{z_-^2}{2}a_{n+2}\,<\,A_n^{\text{Im},\text{log}} \,<\,\\
&&\frac{-1+2\log\frac{z_+}{s}}{2}a_n+\frac{1}{2z_+^2}a_{n-2}.
\label{eq:linearboundsonAimlog}
\eea
While this form might be suited for a numerical implementation in a semidefinite program routine, for a given $\{a_{n-2},a_n,a_{n+2}\}$ one can actually analytically find the optimal $z_\pm$ that give the most stringent constraint. Using these values, one finds
\be
\frac12 \log\frac{a_n}{s^2 a_{n+2}} \,\leq \, \frac{A_n^{\text{Im},\text{log}}}{a_n} \,\leq\, \frac12 \log\frac{a_{n-2}}{s^2 a_{n}}.
\label{eq:boundsonAnIm}
\ee
Note that the homogeneous constraint $a_{n-2}a_{n+2}\geq a_n^2$ ensures that the gap is semipositive. For degenerate values of the arcs $a_{n-2}/s^2\sim a_n\sim s^2 a_{n+2}$, then $A_n^{\text{Im},\text{log}}\sim 0$. Indeed, for degenerate arcs the integral measure is peaked near the threshold at $z\sim s$, and therefore the logarithm vanishes.

In summary, for $n$ even, the integrals over the imaginary and real part of the UV amplitude are written in terms of the IR semiarcs, together with the object $A_n^{\text{Im},\text{log}}$ which is bounded by the IR arcs. The explicit relations are given by
\bea\nn
&& \underline{n\quad \text{even}\quad\quad}\\\nn
A_n^{\text{Im}} &=&  2\text{Re}(a_n^R)\\
A_n^{\text{Re}} &=& -2\text{Im}(a_n^R) -\frac{4}{\pi}\text{Re}(a_n^{R,\text{log}}) + \frac{2}{\pi}A_n^{\text{Im},\text{log}}.
\label{eq:evennrelations}
\eea
While for $n$ odd one finds
\bea\nn
&&  \underline{n\quad \text{odd}\quad\quad}\\\nn
A_n^{\text{Re}} &=&  -2\text{Im}(a_n^R)\\
A_n^{\text{Im}} &=& 2\text{Re}(a_n^R) -\frac{4}{\pi}\text{Im}(a_n^{R,\text{log}}) - \frac{2}{\pi}A_n^{\text{Re},\text{log}}.
\label{eq:oddnrelations}
\eea
The positivity of the imaginary part of the UV amplitude implies that $A_n^{\text{Im}}$ satisfy the Hausdorff moment structure. Moreover, $A_n^{\text{Re}}$ is tied together with $A_n^{\text{Im}}$ via full unitarity.
On the right hand side of the relations, $a_n^R$ and $a_n^{R,\text{log}}$ are calculable within the EFT, while $A_n^{\text{Im},\text{log}}$ can be bounded in terms of the arcs $a_n$.

\smallskip
\textit{Unitarity and null constraints---.}
We discuss now how to find the implications of full unitarity of the 2-to-2 S-matrix to the arcs. 
Using the conventions of \cite{Correia:2020xtr}, the forward amplitude can be written as $\M(s) = \sum_\ell n_\ell f_\ell(s)$, with $n_\ell=16\pi(1+2\ell)$.
In terms of the partial waves $f_\ell$, the unitarity constraint is given by $2\text{Im}f_\ell(s) \geq |f_\ell(s)|^2$ rather than $\text{Im}f_\ell(s)\geq 0$.
We can impose this constraint as it is, giving a nonlinear relation between the imaginary and real parts of the partial waves.
However, one might impose a linear constraint. For $\text{Im}f_\ell$ given, the real part is constrained by
\be
|\text{Re}f_\ell | \leq \frac{1}{\sqrt{x(2-x)}} \left( x +(1-x) \text{Im}f_\ell   \right)
\label{eq:linearunitbound}
\ee
for any $x\in (0,2)$. 
Such linear bounds constrain $A_n^\text{Re}$ in terms of $A_n^\text{Im}$, therefore relating quantities with the same number of subtractions. For $x=1$, one recovers $|\text{Re}f_\ell |\leq 1$, and a similar inequality for $\text{Im}f_\ell$ leads to $2\leq \text{Im}f_\ell\leq 0$.

In its nonlinear form, unitarity relates, in general, semiarcs with different $n$. Using the Cauchy-Schwarz inequality it is possible to bound the integral over $\text{Im}\M(s)$ in terms of the integrals over the real and imaginary parts of the partial waves,
\be
s^n A_n^{\text{Im}} \geq \frac{s^{n+k+1}}{64}\sum_{\ell=0}\frac{1}{1+2\ell}\left[  \left( A_{\frac{n+k-1}{2}}^{\text{Im},\ell} \right)^2 + \left( A_{\frac{n+k-1}{2}}^{\text{Re},\ell} \right)^2 \right],
\label{eq:unitarityCS}
\ee
where $k$ is added so that the number of subtractions $\frac{n+k-1}{2}$ can be an integer, and the label ${}^\ell$ indicates the contribution of only the partial wave $\ell$ to the amplitude.

\smallskip
The key observation that makes Eq.~\ref{eq:unitarityCS} useful is that not all partial waves can give an arbitrary contribution. The reason is that crossing implies non-trivial constraints on the partial waves $f_\ell$. A way to impose crossing is via the null constraints, introduced in Ref.~\cite{Caron-Huot:2020cmc}. The idea behind the null constraints is the following.
One can impose crossing symmetry and polynomial boundedness on $\M(s) = \sum_\ell16\pi(1+2\ell)f_\ell(s)$ in the UV by writing a crossing symmetric ansatz for the amplitude in the IR and imposing the convergence of dispersion relations for enough subtractions. The IR amplitude, at tree level, can be built only from the mass dimension four and six polynomials $X\equiv s^2+t^2+u^2$ and $Y\equiv stu$, so that it is explicitly crossing symmetric. At a given mass dimension $\Delta$, there are $c_\Delta$ independent kinematic invariants of the form $X^iY^j$, where $c_\Delta$ is the coefficient of the generating function $(1-x^4)^{-1}(1-x^6)^{-1}=\sum_\Delta c_\Delta x^\Delta$. 
On the other hand, the $k$-th derivative of the $n$-th arc is in one in one-to-one correspondence with the monomial $s^n t^k$ \footnote{One needs a non-vanishing power of $s^2$ in order to have a convergent dispersion relation. This slightly modifies the counting at low $\Delta$.}.
The number of $s^nt^k$ monomials grows faster than $c_\Delta$, inducing a relation between different $\partial_t^{(k)}a_n$ at each $\Delta$. For instance, at $\Delta=4$ one has that $\partial^{(2)}a_0 \propto a_2$. As a UV relation it translates into a sum rule for the partial waves \cite{Caron-Huot:2020cmc},
\be
A_2^{\text{Im},\ell=2} = \sum_{\ell\geq 4} \frac{\mathcal{J}^2(\mathcal{J}^2-8)}{12} A_2^{\text{Im},\ell},
\label{eq:nullconstraint}
\ee
where we wrote $\mathcal{J}^2\equiv\ell(\ell+1)$.
Given that the weight of the null constraint in Eq.~\ref{eq:nullconstraint} grows as $\sim\ell^4$, in the forward amplitude the $\ell=2$ contribution must completely dominate all the rest of partial waves if one considers enough subtractions.
Studying the arcs and their derivatives for higher $\Delta$ leads to sum rules for higher number of subtractions and higher powers of $\ell$ as well. 
The real parts of the partial waves are bounded by $|\text{Re}f_\ell|\lesssim \sqrt{\text{Im}f_\ell}$ for $\text{Im}f_\ell<< 1$. This might imply slower convergence of the partial waves. 

The above argument holds under the assumption of tree level in the IR. At one loop, the null constraint gets corrected by a logarithm, which might induce modifications to the bounds of certain coefficients \cite{Bellazzini:2021oaj}.
We will focus on qualitative features of EFTs which are under perturbative control, and therefore there is no qualitative difference between the loop and tree-level null constraint. We shall come to this point when discussing bounds on the Wilson coefficients.
At the non-perturbative level, there is no known explicit form for the null constraints. This would be equivalent to solving the crossing and Froissart constraints on the partial wave expansion.

\smallskip
In practice, we rewrite the sum in the right hand side of Eq.~\ref{eq:unitarityCS} as
\be
s^n A_n^{\text{Im}} \geq \frac{s^{n+k+1}}{64}\frac{1}{1+2\ell_{\text{eff}}}\left[  \left( A_{\frac{n+k-1}{2}}^{\text{Im}} \right)^2 + \left( A_{\frac{n+k-1}{2}}^{\text{Re}} \right)^2 \right]
\label{eq:unitarityCSarcs}
\ee
for some $\ell_{\text{eff}}$. Note that the integrals on the right hand side involve a sum over all partial waves. If the amplitude is dominated by a single partial wave, then $\ell_\text{eff}$ coincides with such partial wave. Otherwise, equality between the right hand side of Eq.~\ref{eq:unitarityCS} and Eq.~\ref{eq:unitarityCSarcs} is an equation for $\ell_\text{eff}$. One can write a conservative bound in Eq.~\ref{eq:unitarityCSarcs} by finding the maximal $\ell_{\text{eff}}$ allowed.
The null constraints cut off the contributions from high partial waves, and ensure that such maximal $\ell_\text{eff}$ is finite. For instance, using the null constraint in Eq.~\ref{eq:nullconstraint}, one can show that $\sum_{\ell=0}\frac{1}{1+2\ell}(A^{\text{Im},\ell}_2)^2 \geq \frac{103}{670}(A^{\text{Im}}_2)^2$, corresponding to $\ell_\text{eff}\simeq 2.75$, close indeed to $\ell=2$ as expected.

Using the fact that the sum over partial waves can be truncated for arcs with sufficient subtractions, the linear bounds of Eq.~\ref{eq:linearunitbound} can be used to show that full unitarity forbids arbitrary large hierarchies between arcs. Taking $n$ even, note that both $\text{Im}(a_n^R)$ and $\text{Re}(a_n^{R,\text{log}})$ are finite as long as the amplitude is finite along the semiarc, but the lower bound for $A_n^{\text{Im},\text{log}}$, given by $\frac12 a_n\log\frac{a_n}{s^2 a_{n+2}}$ in Eq.~\ref{eq:boundsonAnIm}, can be arbitrary large for $a_n\gg s^2 a_{n+2}$. Full unitarity constrains such large hierarchies. 
We will come later to this point when considering an explicit realization for the EFT amplitude and give an intuitive argument on why unitarity imposes such constraint. 
We can conclude that the moment structure of the arcs $a_n$ is a necessary but not sufficient self consistency condition of an EFT amplitude.

Alternatively, since both sides of Eq.~\ref{eq:unitarityCSarcs} are calculable, regions of the EFT parameter space can impose a lower bound on $\ell_\text{eff}$, meaning that one cannot UV complete the theory without a sizable contribution from a given $\ell_\text{eff}$. If such required $\ell_\text{eff}$ is larger that the maximal one allowed by the null constraints, such region of parameters space is ruled out.

\section{III.\quad IR implications of UV unitarity}

Unitarity of the UV amplitude imposes a set of relations and constraints on the IR arcs, which in turn translate into a set of constraints on the Wilson coefficients of the EFT.
We consider an EFT of a Goldstone boson, a single massless scalar with a shift symmetry.
Momentarily, we assume the EFT amplitude to be well described by its tree level approximation. In the forward limit, the only consistent structure is a polynomial in the mandelstam variable $s$,
\be
\M(s) \,=\, c_2 s^2 + c_4 s^4 + c_6 s^6 + \dots
\ee
The semiarcs $a_n^R$ can be easily computed, finding that the monomials $s^p$ in the amplitude give a purely imaginary contribution to $a_n^R$, with the notable exception of the case $n=p+2$. 
In this case, there is a one-to-one correspondence between $2\text{Re}(a_n^R)$, or $a_n$, and the Wilson coefficients, $2\text{Re}(a_n^R) = c_{n+2}$.  Therefore, the Wilson coefficients $c_{n+2}=A_n^{\text{Im}}$ are identified with the moments of a positive distribution.
For $n$ odd, both real and imaginary parts of $a_n^R$ receive the leading contribution from $c_2$.

The expressions in Eqns.~\ref{eq:evennrelations} and \ref{eq:oddnrelations} involve the semiarc with a logarithm, $a_n^{R,\text{log}}$,
which receive, in general, the leading contribution from $c_2$, for any $n$.

The fact that the leading contribution to $A_n^{\text{Im}}$ with $n$ even comes from the $c_{n+2}$ Wilson coefficient and $c_2$ is the leading contribution to the rest of the semiarcs is at odds with unitarity. Indeed, the inequality in Eq.~\ref{eq:unitarityCSarcs} implies $s^n c_{n+2} \gtrsim \frac{1}{1+2\ell_{\text{eff}}} \frac{c_2^2}{64\pi^2}$, 
where this should hold up to an order one number from the specific calculation of the right-hand-side. The inequality should be valid for an arbitrary small $c_2$, so the tree level null constraint ensures the convergence of the sum over partial waves for small $\ell_{\text{eff}}$. However, the inequality must hold even in the $s\to 0$ limit, which makes the left-hand-side arbitrary small.
Unitarity implies that the tree-level approximation is inadequate for the calculation of the semiarcs. 

At one loop the forward amplitude receives logarithmic contributions. In the forward limit and for $s$ in the upper half plane, the two loop amplitude is given by \cite{Bellazzini:2020cot}
\bea
\M(s) &=& c_2s^2 + s^4\left[ c_4+\beta_4\log(-is) \right]-i\frac{\pi}{2} s^5\beta_5\\\nn
&+&s^6\left[  c_6 + \beta_6\log(-is) + \beta_6^\prime\log^2(-is)  \right]+\dots,
\eea
where $\beta_4, \beta_5, \dots$ are the beta functions of the Wilson coefficients, calculable within the Goldstone theory in terms of the coefficients $c_2, c_4,\dots$
Such logarithmic correction on the amplitude induces the appearance of contributions proportional to $\beta_4$, etc. to the arcs $A^{\text{Im}}_n$. By dimensional analysis, such contribution is $A^{\text{Im}}_n\sim \beta_4/s^{n-2}$, which is much more relevant than the tree level $A^{\text{Im}}_n\sim c_{n+2}$ contribution, so loop corrections dominate the arcs at low energies for $n\geq 4$.

In fact, as observed in \cite{Bellazzini:2020cot}, positivity of the arcs at all scales within the EFT validity fixes the sign of the most relevant --- in the RG sense--- contribution to the arcs, which is given by $s^n A^{\text{Im}}_n\simeq -\beta_4 + \mathcal{O}(s^2)$. Therefore, the beta function $\beta_4$ must be negative, as it is indeed the case for the Goldstone model.
For $A^{\text{Im}}_2 = 2\text{Re}(a_2^R)$, this contribution is given by the running coefficient $c_4(s)\equiv c_4+\beta_4\log(s)$, so $c_4(s)>0$.

Full unitarity implies not only the negativity of $\beta_4$, but a minimum absolute value.
For the arc with $n=4$, $A^{\text{Im}}_4$, unitarity bounds are of the form $A^{\text{Im}}_4\gtrsim \frac{1}{32}\frac{1}{1+2\ell_\text{eff}} (A^{\text{Re}}_2)^2$. The lower bound also receives contribution from $(A^{\text{Im}}_2)^2$. However, we neglect those because the most relevant contribution to $A^{\text{Im}}_2$ comes from $c_4(s)$, while $A^{\text{Re}}_2$ receives a more relevant contribution $\sim c_2/s^2$.
At arbitrary low energies, using the explicit value for the beta function in the Goldstone theory of $\beta_4=-\frac{7}{10}\frac{c_2^2}{16\pi^2}$, the unitarity bound is given by
\be
\frac{7}{20}\frac{c_2^2}{16\pi^2}\gtrsim \frac{1}{1+2\ell_\text{eff}} \frac{1}{4} \frac{c_2^2}{16\pi^2}
\label{eq:betaconstraint}
\ee
Reassuringly, the explicit value of the beta function does indeed satisfy this bound for $\ell_\text{eff}=0$. This had to be the case, since one can always make $c_2\to 0$, so that the EFT is arbitrarily weakly coupled, and one can UV complete it with just a weakly coupled linear sigma model, so that the most dominant partial wave can be $\ell=0$. 
If Eq.~\ref{eq:betaconstraint} required $\ell_{\text{eff}}\neq 0$ in order to hold, then the UV completion with a Higgs would be inconsistent.
A similar relation is satisfied at any $n$.

Unitarity constraints of the type in Eq.~\ref{eq:unitarityCS} on $A_2^\text{Im}$ impose a lower bound on the running coefficient,
\be
c_4(s)\gtrsim \frac{1}{1+2\ell_{\text{eff}}} \frac{c_2^2}{16\pi^2}.
\ee
Given the negative beta function, this implies that it is not possible to generate an EFT with sizable $c_2$ and vanishingly small $c_4(s)$ when all other contributions to $\text{Re}(a_2^R)$ can be neglected. 

It is worth to understand in a naive language why unitarity imposes such a constraint. In order to have $A^{\text{Im}}_2\ll A^{\text{Im}}_0$, one needs a contribution to the UV integral from a high scale so that the suppression between arcs comes from the extra $1/z^2$ in the integrand. Assuming that such contribution is from some dynamics localized at $z\sim M^2$, with an effective coupling $g$, one has $c_2\sim g^2/M^4$ and $c_4\sim g^2/M^8$. Therefore, in this perturbative setup the bound above implies $g^2\lesssim 16\pi^2$, which is nothing but the naive bound from perturbative unitarity.

Finally, we would like to compare the full unitarity constraints with the ones stemming from the convexity of the moments, or more appropriately from the semipositivity of the Hankel matrix of moments.
At higher energies, $A^{\text{Im}}_4$ gets a tree level contribution from $c_6(s)$. The convexity $A^{\text{Im}}_0A^{\text{Im}}_4\geq (A^{\text{Im}}_2)^2$ of the moments implies the homogeneous constraint
\be
-\frac{\beta_4}{2} + c_6 s^2 \geq \frac{c^2_4(s) s^2}{c_2}.
\label{eq:convexitybetac6}
\ee
In the tree level limit, $\beta_4/(c_6s^2)\ll 1$, this property ensures the convexity of the Wilson coefficients, but the relevant loop correction proportional to $\beta_4$ relaxes this condition. This is to be compared with the constraint from unitarity in Eq.~\ref{eq:betaconstraint} once the $c_6s^2$ term in the arc is included,
\be
-\frac{\beta_4}{2} + c_6 s^2 \gtrsim \frac{1}{1+2\ell_\text{eff}} \frac{1}{4} \frac{c_2^2}{16\pi^2}.
\label{eq:unitaritybetac6}
\ee
Therefore the quantity $-\frac{\beta_4}{2} + c_6 s^2$ receives a constraint from the convexity condition in Eq.~\ref{eq:convexitybetac6} and the unitarity condition in Eq.~\ref{eq:unitaritybetac6}. The importance between the latter and the former is controlled by $\sim \frac{\beta_4}{c_4(s)}\frac{c_2}{c_4(s)s^2}$. The first ratio controls the size of loop corrections, while the second controls the derivative expansion.
If there is no energy regime in which the tree level term $c_6 s^2$ dominates $A^{\text{Im}}_4$, the arc is well described by $\beta_4$ at all scales in the EFT and therefore the naive tree level convexity $c_2c_6\geq c_4^2$ is violated at all scales, and $c_6(s)$ can be negative.
Depending on whether the ratio $\frac{\beta_4}{c_4(s)}\frac{c_2}{c_4(s)s^2}$ is greater or smaller that one, the EFT will reach the cutoff by saturating either the unitarity (Eq.~\ref{eq:unitaritybetac6}) or convexity (Eq.~\ref{eq:convexitybetac6}) constraints.  

In consequence, in theories where loop corrections are more important than higher derivative terms, the unitarity constraints are more important than the ones coming from convexity.

\section{IV. \quad Conclusions}

The semiarcs extend the idea of arcs, and dispersively relate EFT-calculable quantities to the integrals of both the real and imaginary part of the amplitude in the UV for a fixed number of subtractions. 
The relations in Eq.~\ref{eq:evennrelations} and Eq.~\ref{eq:oddnrelations} are the backbone of this \textit{Letter}.

While positivity of the imaginary part of the amplitude implies the Hausdorff moment structure for different subtractions, 
the real part is constrained via full unitarity of the 2-to-2 amplitude in the form of $S^\dagger S\leq 1$. This can be used in the physical, UV energies to derive constraints on the EFT parameters.
We conclude that the moment structure is a necessary but not sufficient consistency condition of the forward scattering amplitude. Large hierarchies between moments are constrained by unitarity.

The null constraints introduced in \cite{Caron-Huot:2020cmc} cut off the contributions from high partial waves, allowing to draw implications of full unitarity for the amplitude.

Loop corrections 
induce a relevant deformation to the arcs that dominates the irrelevant tree level contribution.
Full unitarity provides a lower bound on the absolute value of the beta function of the first running coefficient, and we provide the non-trivial check that the explicit value in a Goldstone model is such that the EFT can indeed be consistently UV-completed by a Higgs.

The version of the full unitarity constraints considered in this \textit{Letter} are similar to the convexity ones, but the former become stronger for EFTs in which the quantum corrections are more important than higher orders of the derivative expansion.

A future study of the semiarcs at finite $t$ would give access to the operators that vanish in the forward limit, as well as the null constraints of the real part of the partial waves. 
At a practical level, it would be interesting to develop a numerical routine that integrates all constraints and extends the simple Cauchy-Schwarz inequalities used in this work. 
A separate question concerns the physical meaning of $I_n$, since it is the quantity controlling the map between unitarity in the UV and its IR implications.

\subsection*{Acknowledgments}

I thank Luc\'ia C\'ordova, Joan Elias Mir\'o, Francesco Riva and Alexander Zhiboedov for discussions and comments on the draft.

\bibliography{bibs} 

\end{document}